\documentclass[times]{elsarticle}

\usepackage[T1]{fontenc}
\usepackage{amsmath}
\usepackage{amsfonts}
\usepackage{amssymb}
\usepackage{graphicx}
\usepackage[dvipsnames]{xcolor}
\usepackage{mmacells}
\usepackage{caption}
\usepackage{subcaption}
\usepackage{enumitem}
\usepackage{float}

\mmaSet{moredefined={LinApart, GaussianInteger, PreCollect, ApplyAfterPreCollect, GatherByDenominator, Parallel}}

\newcommand{\CC}{{\mathbb C}}
\newcommand{\QQ}{{\mathbb Q}}
\newcommand{\ZZ}{{\mathbb Z}}

\def\beq{\begin{equation}}
\def\eeq{\end{equation}}
\def\bsp#1\esp{\begin{split}#1\end{split}}
\def\bal#1\eal{\begin{align}#1\end{align}}

\biboptions{sort&compress}

\begin{document}
\begin{frontmatter}

\title{{\tt LinApart2}: efficient parallel partial fraction decomposition algorithm for denominators with polynomials
of general degree}

\author[1,2]{L.~Fekésházy\corref{cor1}}
\ead{ levente.fekeshazy@desy.de}

\author[1]{O.~Schnetz} 
\ead{oliver.schnetz@mi.uni-erlangen.de}

\cortext[cor1]{Corresponding author}

\affiliation[1]{organization={II.~Institut für Theoretische Physik, Universität Hamburg}, 
	addressline={Luruper Chaussee 149}, 
	postcode={22761}, 
	city={Hamburg}, 
	country={Germany}}
	
\affiliation[2]{organization={Institute for Theoretical Physics, ELTE Eötvös Loránd University}, 
	addressline={P\'azm\'any P\'eter s\'et\'any 1/A}, 
	postcode={1117}, 
	city={Budapest}, 
	country={Hungary}}

\begin{abstract}

We present {\tt LinApart2}, a major update to the {\tt LinApart} algorithm for univariate partial fraction decomposition. Unlike its predecessor, {\tt LinApart2} can handle denominators of arbitrary polynomial degree without explicit factorization, while retaining the efficiency and parallelizability of the Laurent series method. Benchmarks show substantial speedups in both runtime and memory usage compared to {\sc Mathematica}’s built-in routine {\tt Apart} and to the Euclidean algorithm, enabling computations that were previously intractable.
\end{abstract}


\newpageafter{abstract}

\end{frontmatter}

%\iffalse
PROGRAM SUMMARY

\begin{itemize}
\item[]{\em Program title:} {\tt LinApart}

\item[]{\em CPC Library link to program files:}

\item[]{\em Developer's repository link:} \url{https://github.com/fekeshazy/LinApart}

\item[]{\em Licensing provisions:} MIT license (MIT)

\item[]{\em Programming language:} {\sc Wolfram Mathematica}

\item[]{\em Supplementary material:}
\item[]{\em Nature of the problem:}
Analytic computations in Quantum Field Theory frequently yield large expressions involving complex rational functions. In many applications, such as symbolic integration or series expansion, partial fraction decomposition is essential. While univariate partial fraction decomposition is conceptually well-understood and an efficient algorithm exist for linear denominators, there has been no publicly available tool capable of efficiently handling
denominator polynomials of higher degree. Here, we present {\tt LinApart2}, a routine designed to perform univariate partial fraction decomposition in a parallel and highly efficient manner for denominator polynomials of arbitrary degree. Our benchmarks demonstrate significant improvements in timing and memory usage by orders of magnitude for rational expressions with only a few denominator factors. As a result, {\tt LinApart2} enables the decomposition of problems that were previously intractable.

\item[]{\em Solution method:}
The {\tt LinApart2} routine extends the Laurent series method for univariate partial fraction decomposition of rational functions using elements of Galois theory. The {\sc Mathematica} implementation leverages the highly efficient built-in extended polynomial greatest common divider routine and repeated replacement routines to perform the necessary operations.

\item[]{\em Additional comments including Restrictions and Unusual features:}
{\tt LinApart2} preserves the user interface and options of the original version, while introducing parallelization for linear denominators as well.

\item[]{\em References:} \cite{LinApart}
\end{itemize}
\newpage

\section{Introduction}
Partial fraction decomposition (PFD) is a fundamental computational tool in modern perturbative Quantum Field Theory (QFT) calculations, where the increasing complexity of multi-loop and multi-scale problems demands ever more efficient algorithms. Primarily, PFD plays a crucial role in two contexts: simplifying expressions after Integration-By-Parts (IBP) reduction \cite{Feng:2012iq,Bendle:2019csk,Boehm:2020ijp,Badger:2021imn,Badger:2021nhg,Heller:2021qkz}, and preparing integrands for analytic integration in terms of special functions such as multiple polylogarithms \cite{Goncharov:1998kja,Remiddi:1999ew,Goncharov:2001iea,Anastasiou:2013srw,Duhr:2014woa,BrownPar,HyperInt}. In the latter application, partial fraction decomposition is essential for casting integrals into forms recognizable as iterated integrals. Furthermore, direct symbolic computation of phase-space integrals relevant for building subtraction schemes beyond next-to-leading order can lead to expressions where the efficiency of standard partial fraction decomposition routines becomes a bottleneck \cite{Somogyi:2008fc,Bolzoni:2010bt,Somogyi:2013yk,DelDuca:2013kw, NNLOCAL}.
\\

Although univariate partial fraction decomposition is a well-studied problem with numerous algorithmic solutions \cite{Wang81,MAHONEY1983247,Book,kim:2016pf}, the computational demands of contemporary QFT calculations have exposed significant performance limitations of the aforementioned methods. The two most widely implemented approaches are the equation system method and the Euclidean algorithm. However, they suffer from serious scaling problems and offer limited possibility for parallelization. These bottlenecks become particularly acute when dealing with high-degree polynomials, when denominator polynomials are raised to high powers (multiplicities), or when the denominator has many polynomials.

The Laurent series method, introduced in our previous work \cite{LinApart}, addressed these performance concerns by reformulating partial fraction decomposition in terms of a Laurent series calculation. This approach offers two key advantages: first, the residues at different poles can be computed independently, enabling straightforward parallelization; second, the method avoids the polynomial arithmetic bottlenecks in traditional algorithms. However, the original {\tt LinApart} implementation was restricted to cases where the denominator could be completely factored into linear factors over the rationals.
\\

In this paper, we present {\tt LinApart2}, an extension of the Laurent series method that removes this factorization requirement. Our new algorithm performs partial fraction decomposition over irreducible polynomial factors of arbitrary degree, avoiding the introduction of algebraic or complex roots. This is achieved through an application of Galois theory and polynomial reduction techniques, allowing us to express expansion coefficients directly in terms of the polynomial coefficients rather than in terms of polynomial roots.

The key innovation of {\tt LinApart2} lies in its ability to maintain the performance advantages of the Laurent series method while working with unfactored denominators. Our improved algorithm achieves significant speedups compared to traditional methods. Furthermore, the algorithm retains the inherent parallelizability of the Laurent series approach, enabling efficient utilization of modern multi-core architectures.

However, we acknowledge that in some special cases, presented in Section \ref{sec:Performance}, the usage of the Euclidean method or the equation system method proves to be advantageous. Thus, we also include these methods as options in our package.
\\

We provide an open-source implementation of {\tt LinApart2} in the {\sc Wolfram Mathematica} language, designed as a high-performance replacement for the built-in {\tt Apart} command. Benchmarks representing various classes of rational functions demonstrate speedups of several orders of magnitude, with the performance gap widening as the complexity of the problem increases. These improvements enable computations of partial fraction decompositions that were previously intractable due to time or memory constraints.

The remainder of this paper is organized as follows. In Section~\ref{sect:algs}
we define the problem and give an overview of possible algorithms. This section also recaps the Laurent series method for linear denominator polynomials used in {\tt LinApart}.
Section~\ref{EEA} explains the classical method using the extended Euclidean algorithm for denominator polynomials of higher degree. This section ends with a non-trivial, fully worked example.
In Section \ref{GT}, we introduce the Galois method. Applying it to the example of Section \ref{EEA} highlights that the Galois method is conceptually very different from the method
using the Euclidean algorithm.
In Section~\ref{sec:Performance}, we present comprehensive benchmarks comparing {\tt LinApart2} against the built-in {\tt Apart} command and our implementation of the Euclidean algorithm across various classes of rational functions. Finally, Section~\ref{sec:Conclusions} offers conclusions and discusses future extensions of this work.

\section{Algorithms for partial fraction decomposition}
\label{sect:algs}

We consider a rational function in the variable $x$,\footnote{Here and in the following we will denote the decomposition variable by $x$.}
    \begin{gather}
       f(x)=\frac{N(x)}{P(x)}\,,
    \end{gather}
where $N(x)$ and $P(x)$ are polynomials of degrees $\deg N$ and $\deg P$, respectively.

By linearity, we can assume that $N(x)=x^k$ is a power of $x$ although
(depending on the implementation) it can be more efficient to keep the numerator intact.

\subsection{Factorization over the complex numbers}
\label{sectC}
Partial fraction decomposition over the complex numbers $\CC$ requires factorizing $P(x)$ into linear factors, $P(x)=\prod_{i\in I}(x-a_i)^{m_i}$ for some index set $I$.\footnote{Note that a factorization of the denominator into linear factors is beneficial in applications related to symbolic integration in terms of multiple polylogarithms.} The partial fraction decomposition takes the form
    \begin{gather}
    \label{Laurent-series1}
        f(x)
       =
        F(x)+\sum_{i\in I}
            \left(
            \frac{c_{i,1}}{x-a_i} + \frac{c_{i,2}}{(x-a_i)^2} + \cdots 
    	+ \frac{c_{i,m_i}}{(x-a_i)^{m_i}}
        \right)\,,
    \end{gather}
where $F(x)\in\CC[x]$ is the polynomial that corresponds to the expansion at infinity.
This polynomial is determined by a Laurent expansion in $y=1/x$ at $y=0$ to order
$\mathcal{O}(y^0)$. Analogously, we determine the coefficients $c_{i,j}\in\CC$
by expansion at $a_i$,
    \begin{gather}\label{eq:cij_Res}
        c_{i,j}=\frac{1}{(m_i - j)!} \lim_{x\to a_i} \frac{d^{m_i-j}}{dx^{m_i-j}}
    	\Big((x-a_i)^{m_i} f(x)\Big)\,.
    \end{gather}
This formula is straightforward to implement in computer algebra systems. The multiplication by $(x-a_i)^{m_i}$ removes the singularity at $a_i$, reducing the limit to a substitution. Moreover, the independence of the coefficients $c_{i,j}$ (for different $i$) allows parallel computation, offering significant potential for runtime optimization.

The case that all roots $a_i\in\QQ$ are rational is handled in the first version
of {\tt LinApart} using the above expansions.

\subsection{Factorization over a number field}
In many practical applications, factorization of the denominator over $\CC$ is not
practical. Either the zeros are implicit algebraic numbers and only known numerically
(for high $\deg P$) or runtime considerations make working with roots unpractical even for low $\deg P\geq2$.

We therefore consider a factorization of $P(x)$ over a number field
into irreducible polynomials $P_i$ of (possibly) higher degrees,
\begin{equation}\label{Pfact}
    P(x)=\prod_{i\in I} P_i(x)^{m_i}\,.
\end{equation}
Because, typically, one works over the rational numbers, we denote the number field by $\QQ$.
Note, however, that we can replace $\QQ$ with any number field without further changes.

The partial fraction decomposition then takes the form
    \begin{gather}
    \label{Laurent-series2}
        f(x)
       =F(x) +
        \sum_{i\in I}
            \left(
            \frac{N_{i,1}(x)}{P_i(x)} +
            \frac{N_{i,2}(x)}{P_i(x)^2} + \cdots +
            \frac{N_{i,m_1}(x)}{P_i(x)^{m_i}}
        \right)\,,
    \end{gather}
where $F,N_{i,j}\in\QQ[x]$ with $\deg N_{i,j}<\deg P_i$.

Like in the case over $\CC$, the polynomial $F$ can easily be obtained by an expansion at
infinity. So, we will focus on deriving the polynomials $N_{i,j}$.
To do this, there exist three different strategies.
\begin{enumerate}
    \item One can make a general ansatz for the $N_{i,j}$ and bring the right-hand side of \eqref{Laurent-series2} on a common denominator. Comparing coefficients of monomials
    in the numerator with the original expression for $f$ gives a linear system that
    can be solved to determine the coefficients of the $N_{i,j}$.

    This naive approach is implemented in many computer algebra systems
    (also in {\sc Wolfram Mathematica}). Because solving large linear systems
    is very time and memory consuming, this approach has quite catastrophic behavior
    for large expressions.
    \item One can use the extended Euclidean algorithm to reduce
    $|I|$ to one; see Sect.\ \ref{EEA}.
    In the mathematics community, this approach is widely believed to have ideal performance.
    \item One can derive \eqref{Laurent-series2} from \eqref{Laurent-series1} by
    collecting suitable terms over the common denominators $P_i(x)^j$. For this approach,
    it is essential that we only formally (not explicitly) calculate \eqref{Laurent-series1}
    and use elements of Galois theory to derive \eqref{Laurent-series2}; see Sect.\ \ref{GT}.
    
    This method is mathematically more complicated but very competitive in terms of time
    and memory consumption. It is the direct analog of the expansion method for complete
    factorizations into linear factors over $\CC$.    
\end{enumerate}

\section{Partial fraction decomposition using the Euclidean algorithm}
\label{EEA}
\subsection{The Euclidean method}
We assume that the denominator is factorized over $\QQ$ in such a way that
all pairs of factors have greatest common divisor (GCD) $1$. Typically, one completely factors
the denominator over $\QQ$ into irreducible polynomials. This gives the most useful result
for the decomposition.

In the first step, we assume that we have at least two factors $P_1$ and $P_2$ in the denominator.
If this is not the case, we skip the first step.

Because $\gcd(P_1,P_2)=1$ we obtain from the (computationally fast) extended Euclidean algorithm
two polynomials $A_1$ and $A_2$ such that
\begin{equation}\label{eea}
    A_1(x)P_1(x)+A_2(x)P_2(x)=1\quad\text{and}\quad\deg A_1<\deg P_2,\;\deg A_2<\deg P_1\,.
\end{equation}
We multiply the numerator of $f(x)=N(x)/\prod_{i\in I} P_i(x)^{m_i}$ by the left-hand side of
\eqref{eea} to obtain two terms in which one factor in the denominator cancels.
Repeated use of this reduction leads to the case where in each term either $P_1$ or $P_2$ is absent.

We continue by picking the next pair of polynomials in the denominator until the denominator
has only one factor that is raised to some positive integer power.

In the second step,  we assume that $f$ has the form $f(x)=\sum_kN_k(x)/P(x)^k$ for $P=P_i$, $i\in I$
and $1\leq k\leq m_i$. We need to reduce the degrees of the numerators $N_k$ until $\deg N_k<\deg P$.
This is done by polynomial division. We start from $k=m_i$ and obtain
\begin{equation}\label{poldiv}
    \frac{N_{m_i}(x)}{P(x)}=M_{m_i}(x)+\frac{R_{m_i}(x)}{P(x)}\quad\text{with}\quad\deg R_{m_i}(x)<\deg P\,.
\end{equation}
After division by $P(x)^{m_i-1}$, we add $M_{m_i}(x)$ to $N_{m_i-1}(x)$.
We repeat with $k=m_i-1$ and continue until all numerators are reduced.

Alternatively, we can reduce monomials $x^\ell$ for $\ell\geq\deg P$ with polynomial division
and substitute the result until all monomials have degree $<\deg P$.

In the third step, we are left with a collection of polynomials which have to be expanded to
obtain the polynomial $F(x)$ in the partial fraction decomposition \eqref{Laurent-series2}.
In some cases, these terms can become large, so that it is more efficient to ignore all polynomials
obtained in step two and determine $F(x)$ by an expansion at infinity.

\subsection{Example for the Euclidean method}\label{EEM}
\label{exeeq}
Consider the function
\begin{equation}\label{fdef}
    f(x)=\frac{x^{10}}{(x^2+x+1)^2(x^2-x+1)^2}\,.
\end{equation}
The extended Euclidean algorithm for $P_1(x)=x^2+x+1$ and $P_2(x)=x^2-x+1$ yields
\begin{equation}\label{exeeq1}
    \frac{-x+1}2(x^2+x+1)+\frac{x+1}2(x^2-x+1)=1\,.
\end{equation}
We multiply $f(x)$ by the square of the left-hand side (doing two steps at once) and obtain
\begin{equation}
    f(x)=\frac{x^{10}(-x+1)^2}{4(x^2-x+1)^2}+\frac{x^{10}(-x+1)(x+1)}{2(x^2-x+1)(x^2+x+1)}+
    \frac{x^{10}(x+1)^2}{4(x^2+x+1)^2}\,.
\end{equation}

To finish step one, we multiply the middle term with the left-hand side of \eqref{exeeq1} yielding
\begin{equation}\label{eeastep1}
    f(x)=\frac{x^{10}(-x+1)(x+1)^2}{4(x^2+x+1)}+\frac{x^{10}(x+1)^2}{4(x^2+x+1)^2}+\quad(x\leftrightarrow -x)\,.
\end{equation}

In the second step, we use polynomial division to obtain
\begin{equation}
    \frac{x^{10}(x+1)^2}{x^2+x+1}=x^{10}+x^9-x^8+x^6-x^5+x^3-x^2+1+
    \frac{-x-1}{x^2+x+1}\, .
\end{equation}
Substitution into \eqref{eeastep1} gives
\begin{align}
    f(x)=&\frac{-x^{13}-x^{12}+x^{11}+2x^{10}+x^9-x^8+x^6-x^5+x^3-x^2+1}{4(x^2+x+1)}+\frac{-x-1}{4(x^2+x+1)^2}\nonumber\\
    &+\quad(x\leftrightarrow -x)
\end{align}
One last polynomial division yields
\begin{equation}
    \frac{-x^{13}-x^{12}+x^{11}+2x^{10}+x^9-x^8+x^6-x^5+x^3-x^2+1}{x^2+x+1}=
    F_+(x)+\frac{3x + 5}{x^2+x+1}\,,
\end{equation}
where $F_+(x)=-x^{11} + 2x^9 - x^7 + x^5 - 2x^3 + 2x^2 + x - 4$.

Finally, we obtain the polynomial $F$ from $F(x)=(F_+(x)+F_+(-x))/4=x^2-2$. The result is
\begin{equation}\label{pfdex}
    f(x)=x^2-2+\frac{3x+5}{4(x^2+x+1)}+\frac{-x-1}{4(x^2+x+1)^2}+
    \frac{-3x+5}{4(x^2-x+1)}+\frac{x-1}{4(x^2-x+1)^2}\,.
\end{equation}

\section{Partial fraction decomposition using Galois theory}
\label{GT}
\subsection{The Galois method}
In the Galois method, we acquire the polynomial part of the decomposition by expansion at infinity\footnote{The expansion at infinity follows the complex case in Sect.\ \ref{sectC}.}, while we calculate the contribution of each denominator to the Laurent expansion separately. Let us fix an irreducible polynomial $P=P_i$, $i\in I$, in the denominator of $f$ and define
    \begin{equation}
        f(x)=\frac{Q(x)}{P(x)^m} = \frac{Q(x)}{\Big( (x-\alpha_1)\cdots(x-\alpha_n) \Big)^m},
    \end{equation}
where $Q$ is a rational function, $m=m_i$, and $n=\deg P$.
We assume without loss of generality that $Q$ has no factor of $P$ and that $P$ is monic (by absorbing its leading coefficient into $Q$).
\\

As a consequence of irreducibility, $P$ cannot have multiple zeros (otherwise $P$ and $P'$ would have a nontrivial common divisor). Hence, all roots $\alpha_i$ are distinct, and the Galois group of $P$ acts transitively on the $\alpha_i$. This means that all roots
$\alpha_i$ are algebraically equivalent: algebraic equations derived for $\alpha_1$ hold for all $\alpha_i$.

It is important to keep in mind that we work in an algebraic setup. We are not interested in the numerical values of the roots $\alpha_i$ nor in an explicit expression for them. In the following, we only use $P(\alpha_i)=0$.

We fix $\alpha=\alpha_i$ for some $i=1,\ldots,n$ and calculate the pole part of the Lauren series
expansion of $f$ at $\alpha=\alpha_i$. To do this, we do not need the full Galois group of $P$. It suffices to work in the extension $\QQ(\alpha)/\QQ$ of degree $[\QQ(\alpha)/\QQ]=n$. This implies that $\QQ(\alpha)$ is an $n$-dimensional vector space over $\QQ$ with basis $\alpha^k$ for $k=0,\ldots,n-1$,
    \begin{equation}\label{Qbasis}
        \QQ(\alpha)=\QQ\oplus\QQ\alpha\oplus\ldots\oplus\QQ\alpha^{n-1}.
    \end{equation}

We use $P(\alpha)=0$ to work in this basis. Explicitly, this means:
    \begin{itemize}
        \item 
        For any polynomial $N(\alpha)$ we use polynomial division (see \eqref{poldiv}) to obtain
            \begin{equation}
                N(x)=M(x)P(x)+R(x),\quad\deg(R)<n\,.
            \end{equation}
        Therefore, $N(\alpha)=R(\alpha)$, which is in the $\QQ$-basis \eqref{Qbasis}.
        
        In particular $\alpha^n=-a_0-a_1\alpha-\ldots-a_{n-1}\alpha^{n-1}$ if $P(x)=a_0+a_1x+\ldots+a_{n-1}x^{n-1}$. This can be used to iteratively construct tables for monomial reduction.
        \item
        For the inverse of $N(x)$ (which must not be a multiple of $P$) we use the extended Euclidean algorithm \eqref{eea} \footnote{$\gcd(N,P)=1$ because $P$ is irreducible.}
            \begin{equation}
                A(x)N(x)+B(x)P(x)=1,\quad\deg A<\deg P
            \end{equation}
        to obtain $N(\alpha)^{-1}=A(\alpha)$, which is in the $\QQ$-basis \eqref{Qbasis}.
    \end{itemize}

The partial fraction decomposition of $f$ is done in two steps:
    \begin{enumerate}[label=\roman*)]
        \item We calculate the pole part of the Laurent series expansion of $f$ at $x=\alpha$.
        This step is identical to the linear case, with the only extension that the result is in $\QQ(\alpha)$ instead of $Q$.
        We express the result in the basis \eqref{Qbasis} which gives a unique result in terms
        of a list of rational coefficients. This step is identical for all roots $\alpha_i$ of $P$.
        \item We specify $\alpha=\alpha_i$ and sum over $i=1,\ldots,n$. The result is invariant under the Galois group of $P$, which guaranties that we obtain an expression in $\QQ(x)$ where all roots are eliminated.
    \end{enumerate}
In order to perform the actual calculation, we need the following algebraic results:
    \begin{description}
        \item[First step:] \hfill
        
        We define the polynomial
            \begin{equation}\label{Sdef}
                S(x)=\frac{P(x)}{x-\alpha}\in\QQ(\alpha)[x].
            \end{equation}
        Substitution into \eqref{eq:cij_Res} gives
            \begin{equation}\label{aexp}
                f_{\mathrm{pole}}(x)=\sum_{j=1}^m\frac1{(m-j)!}\partial_x^{m-j}\left.\frac{Q(x)}{S(x)^m}\right|_{x=\alpha}\frac1{(x-\alpha)^j}\,,
            \end{equation}
        where $f_{\mathrm{pole}}(x)\in\QQ(\alpha)(x)$ is the pole part of $f$ at $x=\alpha$.
        \\
        
        From \eqref{Sdef} we get
            \begin{equation}\label{Sk}
                P^{(k+1)}(\alpha)=\partial_x^{k+1}(x-\alpha)S(x)\Big|_{x=\alpha}=(k+1)S^{(k)}(\alpha)\,.
            \end{equation}
        Since $Q,P^{(k+1)}\in\QQ(x)$ (not just in $\QQ(\alpha)(x)$), the derivatives in \eqref{aexp} can be worked out in $\QQ(\alpha)$ by using the aforementioned reduction strategies. We obtain an expression of the form
        \begin{equation}\label{aexp1}
            f_{\mathrm{pole}}(x)=\sum_{j=1}^m\frac{b_{j,0}+b_{j,1}\alpha+\ldots+b_{j,n-1}\alpha^{n-1}}{(x-\alpha)^j}
        \end{equation}
        with explicit coefficients $b_{j,k}\in\QQ$.

        \item[Second step:] \hfill

        We set $\alpha=\alpha_i$ and sum over $i$. The coefficients $b_{jk}$ do not depend on $i$, so that by linearity it suffices to determine
            \begin{equation}
                \sum_{i=1}^n\frac{\alpha_i^k}{(x-\alpha_i)^j}
                =\frac{(-1)^{j-1}}{(j-1)!}\partial_x^{j-1}\sum_{i=1}^n\frac{\alpha_i^k}{x-\alpha_i}\,.
            \end{equation}
        By invariance under the Galois group of $P$ the result is in $\QQ(x)$.
        It depends on the coefficients of $P$ but not on the roots $\alpha_i$. Explicitly, we define
            \begin{equation}
                u_k^P(x)=\sum_{i=1}^n\frac{\alpha_i^k}{x-\alpha_i},
            \end{equation}
        for $0\leq k\in\ZZ$. Expanding the fraction by $S$ with $\alpha=\alpha_i$, see \eqref{Sdef}, shows that for every $k$, $u_k^P$ is a rational function in $\QQ(x)$
        with denominator $P$ and numerator degree $\leq n-1$,
            \begin{equation}\label{ukPx}
                u_k^P(x)=\frac{c_{k,0}+c_{k,1}x+\ldots+c_{k,n-1}x^{n-1}}{P(x)},\quad c_{k,0},\ldots,c_{k,n-1}\in\QQ.
            \end{equation}
        The case $k=0$ is
            \begin{equation}
                u_0^P\big(x\big)=\partial_x\log P(x)=\frac{P'(x)}{P(x)}.
            \end{equation}
        The formula for a finite geometric series yields ($\sum_{\ell=0}^1=0$)
            \begin{equation}\label{ured}
                u_k^P(x)=x^ku_0^P(x)-\sum_{i=1}^n\frac{x^k-\alpha_i^k}{x-\alpha_i}=\frac{x^kP'(x)}{P(x)}-\sum_{\ell=1}^kS_{k-\ell}x^{\ell-1},
            \end{equation}
        where
        \begin{equation}
         S\!_j=\sum_{i=1}^n\alpha_i^j.
        \end{equation}
        
        Newton's identities determine the $S\!_j$ in terms of the coefficients of $P$. Explicitly,
            \begin{equation}
                x^{-1}u_0^P(x^{-1})=\sum_{i=1}^n\frac1{1-\alpha_ix}=\sum_{j=0}^\infty x^jS\!_j.
            \end{equation}
        Hence
            \begin{equation}
                S\!_j=\left.\frac1{j!}\partial_x^j\frac{x^{-1}P'(x^{-1})}{P(x^{-1})}\right|_{x=0}.
            \end{equation}
        In fact, Newton's identities give recursive formulas for the $S\!_j$ that may be faster to evaluate than the above differentiation.

        Notice that, by \eqref{ukPx}, many terms in \eqref{ured} cancel out for large $k$.
    \end{description}
With these results, the partial fraction decomposition of $f$ can be calculated.

\subsection{Example for the Galois method}
\label{exgt}
We go back to the example in Sect.\ \ref{EEM}; see Eq.\ \eqref{fdef}.
The expansion at infinity gives
\begin{equation}
    f(x)=\frac{x^2}{(1+x^{-1}+x^{-2})^2(1-x^{-1}+x^{-2})^2}=\frac{x^2}{(1+x^{-2}+x^{-4})^2}
    =x^2-2+\mathcal{O}(x^{-1})\,.
\end{equation}

We set $P(x)=x^2+x+1$, so that $m=2$ and
\begin{equation}
    Q(x)=\frac{x^{10}}{(x^2-x+1)^2}\,.
\end{equation}
Equation \eqref{aexp} gives
\begin{align}
    f_{\mathrm{pole}}(x)&=\partial_x\left.\frac{Q(x)}{S(x)^2}\right|_{x=\alpha}\frac1{x-\alpha}+\frac{Q(\alpha)}{S(\alpha)^2}\frac1{(x-\alpha)^2}\nonumber\\
    &=\left(\frac{Q'(\alpha)}{S(\alpha)^2}-2\frac{Q(\alpha)S'(\alpha)}{S(\alpha)^3}\right)\frac1{x-\alpha}+\frac{Q(\alpha)}{S(\alpha)^2}\frac1{(x-\alpha)^2}\,.
\end{align}
We substitute (see \eqref{Sk})
\begin{equation}
    S(\alpha)=P'(\alpha)=2\alpha+1,\quad S'(\alpha)=\frac12P''(\alpha)=1\,.
\end{equation}
The extended Euclidean algorithm gives
\begin{equation}
    \frac43(x^2+x+1)-\frac{2x+1}3(2x+1)=1,\quad\text{hence}\quad S(\alpha)^{-1}=-\frac{2\alpha+1}3
\end{equation}
and
\begin{equation}
    \frac{-x+1}2(x^2+x+1)+\frac{x+1}2(x^2-x+1)=1,\quad\text{hence}\quad (\alpha^2-\alpha+1)^{-1}=\frac{\alpha+1}2\,.
\end{equation}
Polynomial division gives
\begin{equation}
    x^9=(x^7 - x^6 + x^4 - x^3 + x - 1)(x^2+x+1)+1,\quad\text{hence}\quad \alpha^9=1,\;\alpha^{10}=\alpha\,.
\end{equation}
We obtain
\begin{align}
    Q(\alpha)&=\frac{\alpha(\alpha+1)^2}4=-\frac{\alpha+1}4,\nonumber\\
    Q'(\alpha)&=\frac{10(\alpha+1)^2}4-2\frac{\alpha(\alpha+1)^3}8=\frac{7\alpha-2}4.
\end{align}
Hence
\begin{align}
    f_{\mathrm{pole}}(x)&=\left(\frac{(7\alpha-2)(2\alpha+1)^2}{4\cdot3^2}-\frac{(\alpha+1)(2\alpha+1)^3}{2\cdot3^3}\right)\frac1{x-\alpha}-\frac{(\alpha+1)(2\alpha+1)^2}{4\cdot3^2(x-\alpha)^2}\nonumber\\
    &=\frac{-19\alpha+4}{36(x-\alpha)}+\frac{\alpha+1}{12(x-\alpha)^2}\,.
\end{align}
We have (see \eqref{ured})
\begin{align}
    -19u^P_1(x)+4u^P_0(x)&=\frac{(-19x+4)(2x+1)}{P(x)}+38=3\frac{9x+14}{P(x)}\,,\nonumber\\
    u^P_1(x)+u^P_0(x)&=\frac{(x+1)(2x+1)}{P(x)}-2=\frac{x-1}{P(x)}\,.
\end{align}
Hence,
\begin{align}
    \sum_{i=1}^2\frac{-19\alpha_i+4}{36(x-\alpha_i)}&=\frac{9x+14}{12P(x)}\,,\nonumber\\
    \sum_{i=1}^2\frac{\alpha_i+1}{12(x-\alpha_i)^2}&=-\partial_x\frac{x-1}{12P(x)}=-\frac1{12P(x)}+
    \frac{(x-1)(2x+1)}{12P(x)^2}=\frac1{12P(x)}-\frac{x+1}{4P(x)^2}\,.
\end{align}
Altogether, we obtain
\begin{equation}
    f_{\mathrm{pole}}(x)=\frac{3x+5}{4P(x)}-\frac{x+1}{4P(x)^2}
\end{equation}
for the pole part for $P(x)$. The pole part for $P(-x)$ is given by reflection symmetry.
We reproduce \eqref{pfdex} for the partial fraction decomposition of $f$.

In this example, the Galois method is more complicated than the Euclidean method.
However, we will see from our benchmarks, Sect. \ref{sec:Performance}, that the Galois method
scales better with more factors in the denominator. It is also particularly suited for parallelization.

\subsection{Parameters}
In the previous sections, we worked over a number field (for explicitness, we used $\QQ$).
In applications, the polynomials may have coefficients that are parameters $c_i$.
In this case, the number field $\QQ$ is replaced by the function field $\QQ(c_i)$.
In this function field, the algorithm is formally the same as in $\QQ$ (with a considerable increase in time and memory consumption). Most of the benchmarks in Sect.\ \ref{sec:Performance} are in a setup where all coefficients are parameters.

\section{Performance}
\label{sec:Performance}

The algorithms presented above have language specific implementations. Their performance is strongly tied to the computer algebra environment in which they are written. For example, in FORM \cite{FORM} the Euclidean algorithm has a clear advantage due to FORM's ability to quickly expand expressions and apply substitution rules. In {\sc Wolfram Mathematica} we can leverage the speed of the built-in derivative function, like we did in the first version of {\tt LinApart}. Out of convenience and preference, we have constrained ourselves to the {\sc Wolfram Mathematica} language. 
\\

In this section, we will present a comprehensive but not all-encompassing comparison between the Euclidean and Galois method. During our benchmarks we measured the time and memory need as a function of various ``complexities''.

In general, a rational fraction's complexity can come from several factors, for example from:
    \begin{enumerate}[label=(\roman*)]
        \item the number of distinct denominator factors.
        \item the degree of each individual denominator.
        \item the algebraic complexity of the polynomial coefficients of the denominators.
        \item the multiplicity (the exponent) of the denominators.
        \item the degree of the numerator.
    \end{enumerate}.
\\

Since each algorithm exhibits different sensitivities to these factors, we varied one complexity parameter at a time (subject to resource constraints) to isolate their effects. We restrict ourselves to decomposition computations that are completed in an hour using less than 16 GB of RAM. We think these limits reflect practical constraints and typical real-world usage. In some figures the built-in function {\tt Apart} is excluded, since it could not provide results.
\\

In our previous work \cite{LinApart}, we examined the effect of increasing the number of linear denominators. With the new algorithm, we are able to compare our method against the Euclidean method also in the case of nonlinear, irreducible denominators.

\begin{figure}[!htbp]
\vspace*{-1cm}
%––––– first pair ––––––––––––––––––––––––––––––––––––––
    \begin{subfigure}[t]{\textwidth}
        \hspace*{-2.0cm}
        \centering
        \begin{minipage}{0.5\textwidth}
            \centering
            \includegraphics[scale=1.5]{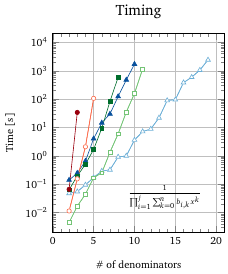}
        \end{minipage}%
        \begin{minipage}{0.5\textwidth}
            \centering
            \includegraphics[scale=1.5]{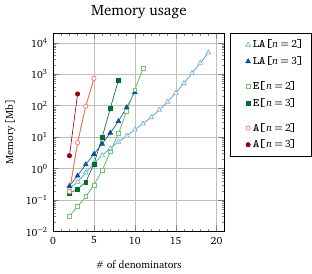}
        \end{minipage}
        \caption{\label{subfig:plot_for_num_of_denoms} Benchmark for increasing number of second and third order denominators with multiplicity one. The lines with hollow symbols denote the case of quadratic denominators, while the solid symbols denote the case when the denominators are cubic.}
    \end{subfigure}

    \par\vspace{1.5cm}     % vertical gap

%––––– second pair ––––––––––––––––––––––––––––––––––––––
    \begin{subfigure}[t]{\textwidth}
        \hspace*{-2.0cm}
        \centering
        \begin{minipage}{0.5\textwidth}
            \centering
            \includegraphics[scale=1.5]{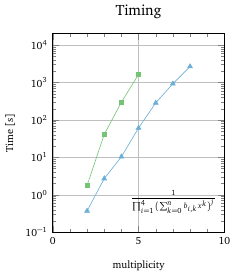}
        \end{minipage}%
        \begin{minipage}{0.5\textwidth}
            \centering
            \includegraphics[scale=1.5]{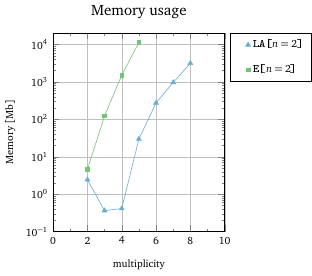}
        \end{minipage}
        \caption{\label{subfig:plot_for_multiplicity} Benchmark for increasing the multiplicity of a fixed number of second order denominators. In this case, we used four denominators and increased the multiplicity of all of them.}
    \end{subfigure}
    
\caption{\label{fig:Benchmarks1} Timings and memory usage of the new {\tt LinApart} function, {\tt Apart} and the Euclidean algorithm (denoted as \texttt{LA}, \texttt{A} and \texttt{E} in the legend) in case of different rational functions with symbolic polynomial coefficients. In (a) we plotted the benchmarks with increasing number of denominators with different polynomial degree ($n$), while in (b) we show the same metrics for a fixed number of denominators with increasing multiplicity.}

\end{figure}

In these benchmarks, we computed the partial fraction decomposition with respect to $x$ of rational functions of the form
    \begin{gather}
    f(x) 
   =
    \frac{1}{\prod_{i=1}^{j} \sum_{k=0}^{n} b_{i,k}\,x^k},
    \end{gather}
where $j$ was varied and $n \in \{2,3\}$. The time and memory usage are shown in Figure \ref{subfig:plot_for_num_of_denoms}. It is evident that {\tt LinApart2} significantly outperforms both {\tt Apart} and the Euclidean algorithm in terms of runtime and memory consumption. In particular, for $n=2$ and $j=10$, {\tt LinApart2} is nearly two orders of magnitude faster than the Euclidean method. At the same time, the plots also reveal that {\tt LinApart2} is more sensitive to the degree of the denominators than the Euclidean method or {\tt Apart}. According to our benchmarks, this behavior stems from the speed of polynomial reduction, which constitutes the main bottleneck.
\\

Next, we considered the case of a fixed number of denominators whose multiplicities were increased simultaneously:
    \begin{gather}
    f(x) 
   =
    \frac{1}{\prod_{i=1}^{4} \big( \sum_{k=0}^{n} b_{i,k}\,x^k \big)^j},
    \end{gather}
for various $j$ and $n=2$. The results are shown in Figure~\ref{subfig:plot_for_multiplicity}. Here, {\tt Apart} was unable to process even the simplest examples, and therefore its metrics could not be included. Comparing the two remaining methods, {\tt LinApart2} exhibits a clear advantage, especially in terms of memory consumption, where it outperforms the Euclidean algorithm by at least two orders of magnitude.
\\

\begin{figure}[!htbp]
\vspace*{-2.5cm}

%––––– first pair ––––––––––––––––––––––––––––––––––––––
    \begin{subfigure}[t]{\textwidth}
        \hspace*{-2.0cm}
        \centering
        \begin{minipage}{0.5\textwidth}
            \centering
            \includegraphics[scale=1.5]{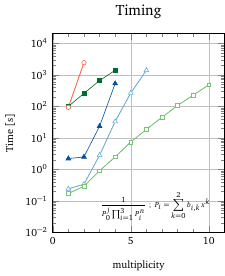}
        \end{minipage}%
        \begin{minipage}{0.5\textwidth}
            \centering
            \includegraphics[scale=1.5]{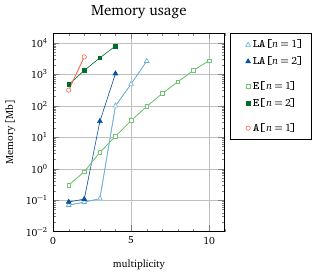}
        \end{minipage}
        \caption{\label{subfig:plot_for_one_multiplicity} Benchmark for increasing the multiplicity of a fix number of second order denominators. In this case we used four denominators and increased the multiplicity of only one. The lines with hollow symbols denote the case when the auxiliary denominators have a multiplicity of $n=1$, while the solid symbols denote the case of $n=2$.}
    \end{subfigure}

    \par\vspace{1.5cm}     % vertical gap

%––––– second pair ––––––––––––––––––––––––––––––––––––––
    \begin{subfigure}[t]{\textwidth}
        \hspace*{-2.0cm}
        \centering
        \begin{minipage}{0.5\textwidth}
            \centering
            \includegraphics[scale=1.5]{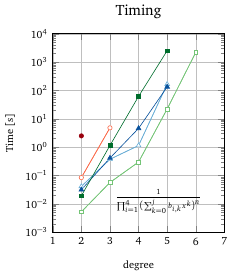}
        \end{minipage}%
        \begin{minipage}{0.5\textwidth}
            \centering
            \includegraphics[scale=1.5]{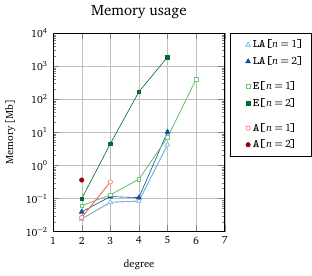}
        \end{minipage}
        \caption{\label{subfig:plot_for_order} Benchmark for a fixed number of denominators with increasing degree. The lines with hollow symbols denote the case of multiplicity one, while the solid symbols denote the case when the denominators are squared.}
    \end{subfigure}

\caption{\label{fig:Benchmarks2} Timings and memory usage of the new {\tt LinApart} function, {\tt Apart} and the Euclidean algorithm (denoted as \texttt{LA}, \texttt{A} and \texttt{E} in the legend) on different rational functions with symbolic polynomial coefficients. In (a) we plotted the benchmarks where we increase the multiplicity of only one denominator; the denominators had different polynomial degree ($n$). In (b) we show the aforementioned metrics for a fixed number of denominators with increasing degree. Here $n$ denotes the multiplicity.}

\end{figure}

In the following, we fixed the number of denominators and increased the multiplicity of only one of them. In this rather simple case we have
    \begin{gather}
    f(x) =
        \frac{1}{(b_{0,0}+b_{0,1}x+b_{0,2}x^2)^j}\,\frac{1}{\prod_{i=1}^{3} \big( \sum_{k=0}^{2} b_{i,k}\,x^k \big)^n},
    \end{gather}
for various $j$ and $n \in \{1,2\}$. As Figure \ref{subfig:plot_for_one_multiplicity} shows, for simple poles the Euclidean algorithm is orders of magnitude faster than either {\tt LinApart2} or {\tt Apart}. Although {\tt LinApart2} is competitive for low multiplicities, increasing the multiplicities of the other denominators does not improve its scaling. This suggests, given larger time and memory limits, that the Euclidean algorithm would remain more efficient in the high-multiplicity regime in essentially all cases.
\\

We also investigated the effect of increasing the polynomial degree of the denominators for a fixed number of denominators:
    \begin{gather}
    f(x) 
   =
    \frac{1}{\prod_{i=1}^{4} \big( \sum_{k=0}^{j} b_{i,k}\,x^k \big)^n},
    \end{gather}
for various $j$ and $n \in \{1,2\}$. As one can see on Figure \ref{subfig:plot_for_order}, in this benchmark the situation is reversed compared to the previous case; while the Euclidean algorithm is initially faster, {\tt LinApart2} exhibits much better scaling, especially in terms of memory consumption.

\hfill

Finally, we benchmarked the case of non-symbolic coefficients by substituting random integers. In this regime {\tt LinApart2} does not dominate every benchmark; nevertheless, its relative performance improves substantially as complexity increases. This is to be expected: the equation-based method has an advantage when only a single symbol is present, but as soon as additional symbols enter the coefficients this advantage diminishes. For example, with only one variable

\begin{mmaCell}{Code}
\mmaDef{expr}=x^2/\mmaDef{Product}[
                Sum[
                    \mmaDef{RandomInteger}[{1,10^5}] x^j,
                    {j,0,3}
                ]^2,
                {i,1,5}
            ];

\mmaDef{expr}//\mmaDef{LinApart}[#,x]&;//\mmaDef{AbsoluteTiming}
\mmaDef{expr}//\mmaDef{Apart}[#,x]&;//\mmaDef{AbsoluteTiming}

\end{mmaCell}

\begin{mmaCell}{Output}
\{0.588255, \mmaDef{Null}\}
\end{mmaCell}

\begin{mmaCell}{Output}
\{0.044992, \mmaDef{Null}\}
\end{mmaCell}

but when we introduce additional symbolic parameters:

\begin{mmaCell}{Code}
\mmaDef{expr}=x^2/\mmaDef{Product}[
                Sum[
                    (\mmaDef{RandomInteger}[{1,10^5}] y +
                        \mmaDef{RandomInteger}[{1,10^5}] a) x^j,
                    {j,0,3}
                ]^2,
                {i,1,5}
            ];

\mmaDef{expr}//\mmaDef{LinApart}[#,x]&;//\mmaDef{AbsoluteTiming}
\mmaDef{expr}//\mmaDef{Apart}[#,x]&;//\mmaDef{AbsoluteTiming}

\end{mmaCell}

\begin{mmaCell}{Output}
\{3.73804, \mmaDef{Null}\}
\end{mmaCell}

\begin{mmaCell}{Output}
\{140.521, \mmaDef{Null}\}
\end{mmaCell}
\begin{figure}[!htbp]

%––––– first pair ––––––––––––––––––––––––––––––––––––––
    \begin{subfigure}[t]{\textwidth}
        \hspace*{-2.5cm}
        \centering
        \begin{minipage}{0.5\textwidth}
            \centering
            \includegraphics[scale=1.5]{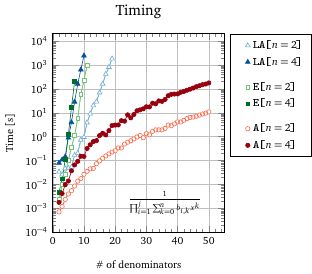}
        \end{minipage}%
        \hspace{2.0cm}
        \begin{minipage}{0.5\textwidth}
            \centering
            \includegraphics[scale=1.5]{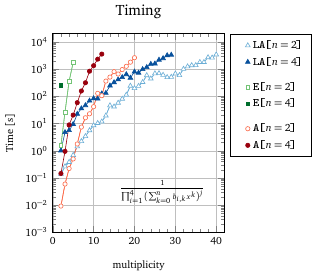}
        \end{minipage}
    \end{subfigure}

    \par\vspace{1.5cm}     % vertical gap

%––––– second pair ––––––––––––––––––––––––––––––––––––––
    \begin{subfigure}[t]{\textwidth}
        \hspace*{-2.5cm}
        \centering
        \begin{minipage}{0.5\textwidth}
            \centering
            \includegraphics[scale=1.5]{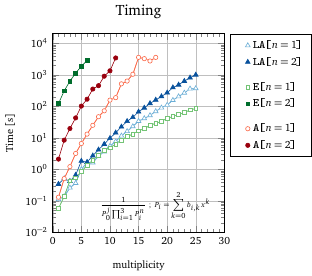}
        \end{minipage}%
        \hspace{2.0cm}
        \begin{minipage}{0.5\textwidth}
            \centering
            \includegraphics[scale=1.5]{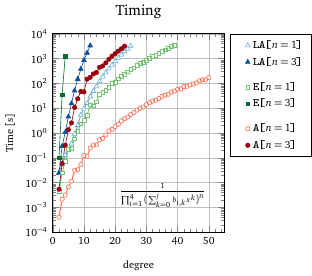}
        \end{minipage}
    \end{subfigure}
    
    \caption{\label{fig:plot_for_numerical_benchmarks} Different benchmarks in the case of integer polynomial coefficients. The benchmarked cases are the same as in the previous fully symbolic coefficient case in Figures \ref{fig:Benchmarks1} and Figures \ref{fig:Benchmarks2}.}
\end{figure}

One of the main advantages and the pivot point of our algorithm is the possibility of parallelization. We provide this as a new option to the {\tt LinApart} function in the following way\footnotemark. 

\footnotetext{For parallelization we have followed the strategy outlined in this blog post \url{https://community.wolfram.com/groups/-/m/t/3540598?p_p_auth=zb1K16fL}.}

\begin{table}[H]
    \centering
        \renewcommand{\arraystretch}{1.5}
        \begin{tabular}{|p{0.32\textwidth}|p{0.58\textwidth}|}
        \hline 
        New Option & Description \\
        \hline\hline
        {\tt Parallel} & 
        { \footnotesize
        \{{\tt True}/{\tt False},  {\tt NumberOfCores\_Integer},  {\tt \$PATH\_String}\}: If the first argument is set to {\tt True} parallelization is enabled and the user must give the number of cores, which the algorithm may use, and a path to the temporary files. 
        }
        \\
        \hline
        {\tt Extension} & 
        { \footnotesize
        \{$a_1, a_2, ...$\}: option for Factor; factors a polynomial allowing coefficients that \
are rational combinations of the algebraic numbers $a_i$.
        }
        \\
        \hline
    \end{tabular}
\end{table}

However, the effectiveness of such a routine is highly language dependent, as each system treats sub-workers (or subkernels) and memory management differently. In {\sc Wolfram Mathematica}, parallelization regarding symbolic computations is notoriously challenging. Unfortunately, {\sc Mathematica}'s memory management prevents us from gaining significant leverage over the single-core computations in every case. But we would like to emphasize, that this issue is solely due to {\sc Mathematica}'s restrictions and does not originate from our algorithm. Hence, in other languages, which allow the user to influence memory management on a deeper level, they should not arise.
\\

In the linear case, this can be illustrated with two simple examples. For many denominators with simple poles, parallelization in {\sc Mathematica} provides essentially no speed-up.

\begin{mmaCell}{Code}
\mmaDef{expr}=x^2/\mmaDef{Product}[\mmaDef{Sum}[b[i,j] x^j, {j,0,1}], {i,1,30}];

\mmaDef{expr}//\mmaDef{LinApart}[#,x]&;//\mmaDef{AbsoluteTiming}
\mmaDef{expr}//\mmaDef{LinApart}[
                #,x,
                "Parallel"->{True,4,\mmaDef{NotebookDirectory}[]}
            ]&;//\mmaDef{AbsoluteTiming}

\end{mmaCell}

\begin{mmaCell}{Output}
\{24.0016, \mmaDef{Null}\}
\end{mmaCell}

\begin{mmaCell}{Output}
\{21.1083, \mmaDef{Null}\}
\end{mmaCell}

However, in the high-multiplicity limit, runtime can indeed be reduced:

\begin{mmaCell}{Code}
\mmaDef{expr}=x^2/\mmaDef{Product}[(x-b[i,1])^20, {i,1,15}];

\mmaDef{expr}//\mmaDef{LinApart}[#,x]&;//\mmaDef{AbsoluteTiming}
\mmaDef{expr}//\mmaDef{LinApart}[
                #,x,
                "Parallel"->{True,4,\mmaDef{NotebookDirectory}[]}
            ]&;//\mmaDef{AbsoluteTiming}

\end{mmaCell}

\begin{mmaCell}{Output}
\{29.8126, \mmaDef{Null}\}
\end{mmaCell}

\begin{mmaCell}{Output}
\{13.6825, \mmaDef{Null}\}
\end{mmaCell}

In the general case, the situation is similar. If the denominators are simple or of low multiplicity but high degree, parallelization again offers no speed-up:

\begin{mmaCell}{Code}
\mmaDef{expr}=x^2/\mmaDef{Product}[\mmaDef{Sum}[b[i,j] x^j, {j,0,2}]^2, {i,1,8}];

\mmaDef{expr}//\mmaDef{LinApart}[#,x]&;//\mmaDef{AbsoluteTiming}
\mmaDef{expr}//\mmaDef{LinApart}[
                #,x,
                "Parallel"->{True,4,\mmaDef{NotebookDirectory}[]}
            ]&;//\mmaDef{AbsoluteTiming}
            
\end{mmaCell}

\begin{mmaCell}{Output}
\{31.7012, \mmaDef{Null}\}
\end{mmaCell}

\begin{mmaCell}{Output}
\{30.7866, \mmaDef{Null}\}
\end{mmaCell}

Nevertheless, it still offers advantages in specific scenarios; namely, it reduces runtime for high multiplicity cases:

\begin{mmaCell}{Code}
\mmaDef{expr}=x^2/\mmaDef{Product}[\mmaDef{Sum}[b[i,j] x^j, {j,0,2}]^4, {i,1,5}];

\mmaDef{expr}//\mmaDef{LinApart}[#,x]&;//\mmaDef{AbsoluteTiming}
\mmaDef{expr}//\mmaDef{LinApart}[
                #,x,
                "Parallel"->{True,4,\mmaDef{NotebookDirectory}[]}
            ]&;//\mmaDef{AbsoluteTiming}
            
\end{mmaCell}

\begin{mmaCell}{Output}
\{148.413, \mmaDef{Null}\}
\end{mmaCell}

\begin{mmaCell}{Output}
\{103.241, \mmaDef{Null}\}
\end{mmaCell}

The poorest performance occurs when the degrees of the denominators are increased. 

\begin{mmaCell}{Code}
\mmaDef{expr}=x^2/\mmaDef{Product}[\mmaDef{Sum}[b[i,j] x^j, {j,0,2}]^2, {i,1,7}];

\mmaDef{expr}//\mmaDef{LinApart}[#,x]&;//\mmaDef{AbsoluteTiming}
\mmaDef{expr}//\mmaDef{LinApart}[
                #,x,
                "Parallel"->{True,4,\mmaDef{NotebookDirectory}[]}
            ]&;//\mmaDef{AbsoluteTiming}
            
\end{mmaCell}

\begin{mmaCell}{Output}
\{7.02244, \mmaDef{Null}\}
\end{mmaCell}

\begin{mmaCell}{Output}
\{5.5188, \mmaDef{Null}\}
\end{mmaCell}

\begin{mmaCell}{Code}
\mmaDef{expr}=x^2/\mmaDef{Product}[\mmaDef{Sum}[b[i,j] x^j, {j,0,3}]^2, {i,1,5}];

\mmaDef{expr}//\mmaDef{LinApart}[#,x]&;//\mmaDef{AbsoluteTiming}
\mmaDef{expr}//\mmaDef{LinApart}[
                #,x,
                "Parallel"->{True,4,\mmaDef{NotebookDirectory}[]}
            ]&;//\mmaDef{AbsoluteTiming}
            
\end{mmaCell}

\begin{mmaCell}{Output}
\{8.86809, \mmaDef{Null}\}
\end{mmaCell}

\begin{mmaCell}{Output}
\{13.8237, \mmaDef{Null}\}
\end{mmaCell}

The bottleneck lies in {\sc Mathematica}’s memory management: after the subkernels complete their tasks, the entire expression must be copied back to the main kernel, which takes significant time. Furthermore, since the main kernel encounters this expression for the first time, it was never cached. Thus, combining the partial results is slower than if the computation had taken place entirely in the main kernel.

\section{Conclusions and outlook}
\label{sec:Conclusions}

In this article, we have presented a new extended version of the {\tt LinApart} algorithm; a significant advancement compared to the original algorithm in partial fraction decomposition. This new method does not require explicit factorization of the denominator, thus overcoming the main usage limitation of the original algorithm
in case of denominator polynomials of higher degree. The new approach is particularly advantageous when dealing with polynomial denominators whose roots involve algebraic or complex numbers, or where runtime considerations make complete factorization impractical.

We showcased the capabilities of our algorithm with the help of a {\sc Wolfram Mathematica} implementation, which can be used out-of-the-box and is capable of completely replacing the built-in {\tt Apart} function. We have thoroughly benchmarked our implementation, comparing its timings against both {\sc Mathematica}'s {\tt Apart} function and our own implementation of the method using the Euclidean algorithm. The figures show that our algorithm is advantageous in most use-cases, only underperforming slightly against the Euclidean-algorithm in one special case.

Of course, the runtime and memory benchmarks do not only depend on the choice of {\sc Wolfram Mathematica} as computer algebra system, but also on our specific implementation. It is possible that in future versions, improvements can be found in the implementation of either the Euclidean method or the Galois method that shift the line between which method is advantageous in which setup. From our experience, backed by the benchmarks presented in Section \ref{sec:Performance}, the Eucliden method or even {\tt Apart} tends to be more efficient in the simplest setups, whereas {\tt LinApart} proves to be more powerful as the complexity of the problem increases.
\\

However, we would like to emphasize that {\tt LinApart2} is the only partial fraction decomposition algorithm which offers a way of efficient parallelization, opening the possibility of further run-time optimization. Although the shortcomings of {\sc Wolfram Mathematica} prevented us from fully utilizing this potential, we could still observe significant run-time improvements in some cases, serving as a compelling proof-of-concept. As this issue is language specific and would be absent in other computer algebra languages, we claim that the {\tt LinApart2} algorithm is not just an incremental improvement, but a prime candidate for an efficient all-around single variable partial decomposition algorithm. 
\\

Lastly, we note that, the mathematical considerations in Sect.\ \ref{sect:algs}--\ref{GT} are entirely self-contained. They may serve as a basis for future implementation in other computer algebra systems.

\section*{Acknowledgements}

The authors would like to thank Prof. Dr. Sven-Olaf Moch for his comments on the manuscript. The work of L.F. was supported by the German Academic Exchange Service (DAAD) through its Bi-Nationally Supervised Scholarship program. The work of O.S. is financed by the DFG grant SCHN 1240/3-1.

\bibliographystyle{elsarticle-num}
\bibliography{omebib}

\end{document}